\documentclass[12pt]{revtex4}
\usepackage{graphicx, color}
\begin{document}

\title[Spin melting and refreezing driven by uniaxial compression on a dipolar hexagonal plate]{Spin melting and refreezing driven by uniaxial compression on a dipolar hexagonal plate
}
\author{K Matsushita$^1$, R Sugano$^2$, A Kuroda$^3$, Y Tomita$^3$ and H Takayama$^3$}

\date{empty}

\address{$^1$ National Institute for Materials Science, 1-2-1 Sengen, Tsukuba, Ibaraki 305-0047, Japan}
\address{$^2$ Advanced Research Laboratory, Hitachi, Ltd.,
  1-280 Higashi-Koigakubo, Kokubunji-shi, Tokyo 185-8601, Japan}
\address{$^3$ Institute for Solid State Physics, University of
  Tokyo, 5-1-5 Kashiwanoha, Kashiwa, Chiba 277-8581, Japan}
\email{MATSUSHITA.Katsuyoshi@nims.go.jp}
\begin{abstract}
We investigate freezing characteristics of a finite dipolar hexagonal
 plate by the Monte Carlo simulation. 
The hexagonal plate is cut out from 
a piled triangular lattice of three layers 
with FCC-like (ABCABC) stacking structure. 
In the present study an annealing simulation is performed for the
 dipolar plate uniaxially compressed in the direction of layer-piling.  
We find spin melting and refreezing driven by the 
uniaxial compression. 
Each of the melting and refreezing corresponds one-to-one with a change 
of the ground states induced by compression.
The freezing temperatures of the ground-state orders differ
 significantly from each other, which gives rise to the spin melting and
 refreezing of the present interest.
We argue that these phenomena are originated by a finite size effect
 combined with peculiar anisotropic nature of the dipole-dipole
 interaction.
 
\end{abstract}

\maketitle

\section{Introduction}

In recent years, systems consisting of arrayed single-domain
ferromagnetic nanoparticles receive much attention of many researchers
because of its possibility as a high storage density~\cite{Martin}. 
In such array systems the dipole-dipole interaction 
is considered to play
an important role on their magnetic properties. 
The interaction not only intrinsically involves frustration, i.e.,
the competition between ferromagnetic and antiferromagnetic
interactions, but also it connects the anisotropy in the the spin space
with that in the real space. 
In finite systems the latter is expected to yield a vast variety of
ground-state spin configurations depending on shapes of the systems.
Various ordered spin patterns have been actually observed, for example,
in infinite~\cite{Garel,Igresias,Jagla} or
finite~\cite{Belobrov,Vedmedenko} systems only with the dipole-dipole
interaction, which we call {\it dipolar systems}, on the triangular
lattice.  
In dipolar systems such a peculiar phenomenon as the  
from-Edge-to-Interior freezing has been also found 
recently~\cite{KSKTT}. 
In the present study we investigate a dipolar hexagonal plate cut out
from a piled triangular lattice of three layers with FCC-like structures
and with uniaxial compression in the direction of layer-piling. 
Our aim here is to demonstrate that, in the dipolar system even with
such a restricted shape, an interesting crossover is predicted to exist
when the interlayer distance $l$ is changed.

\begin{figure}[b]
\begin{center}
\includegraphics[width=0.75\linewidth]{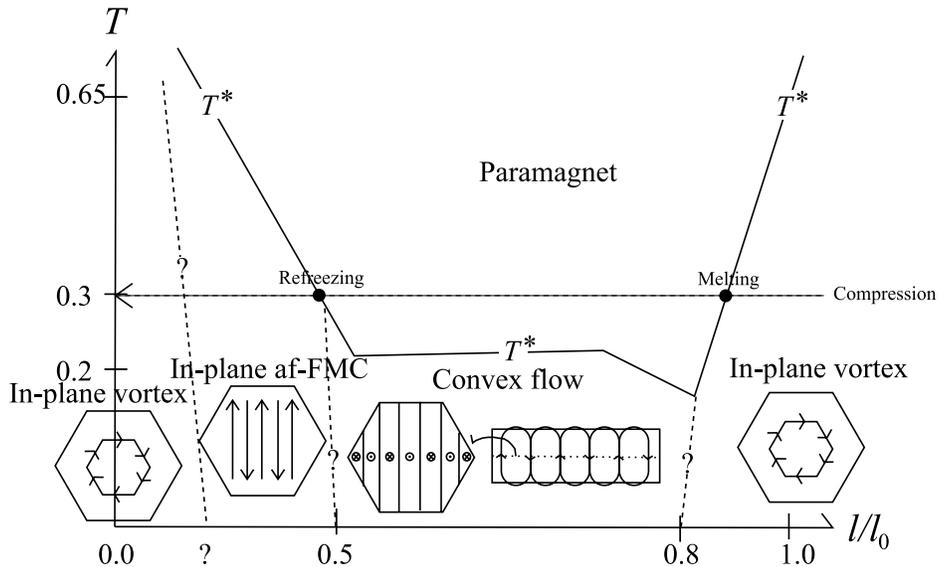}
\end{center}
\caption{The schematic 
spin-pattern and $T^*$
diagram of the dipolar hexagonal plate. 
Arrows in insets shows directions of spins. }\label{SPD}
\end{figure} 

Figure \ref{SPD} shows a schematic spin-pattern and $T^*$ diagram obtained
by the present study of the dipolar hexagonal 
plate, where $T^*$ is the temperature at which spins start to be frozen. 
At low temperatures there appear three kinds of patterns, i.e.,   
in-plane-vortex, convex-flow, in-plane-af-FMC (see Fig.~\ref{GS}(c)
below) patterns depending on a value of $l$. 
The freezing temperature, $T^*$, depends on a spin pattern of the
corresponding ground state.
Especially, $T^*$ of the convex flow state is much lower than 
those of other states. 
Because of this $l$-dependence of $T^*$, we expect the
peculiar spin melting and refreezing driven by uniaxial 
compression, say, at $T=0.3$.

In the present study we perform simulated annealing of the dipolar 
hexagonal plate by the Monte Carlo (MC) method. 
The well-known Hamiltonian of a dipolar system,
\begin{center}
\begin{eqnarray}
H = \frac{D}{2a^3}\sum_{<i,j>} {\vec S}_i \cdot \frac{1-{\vec e}_{ij} \otimes {\vec e}_{ij} }{r_{ij}^3} \cdot {\vec S}_j, \label{Ham}
\end{eqnarray}
\end{center}
is employed.
Here $\vec S_i$'s denote classical Heisenberg spins which represent
magnetic moments of nanoparticles, $D$ a coupling constant, and $a$ a
lattice constant of the triangular lattice.
The number of piled layers in a plate examined, $N_{\rm L}$, 
is 3. 
Each layer contains about 100 sites.
The $z$-direction is set in perpendicular to the layers.  
We perform typically 8 cooling runs with a fixed cooling rate, in which 
we use different sets of random numbers generating their initial
configurations as well as spin-renewing for each MC step. 
For each cooling run, temperature $T$ is initially set to about 1 in
unit of $D/a^3$ and is decreased by a step of $\Delta T = 0.0025$. 
At each temperature, states are sampled over $\tau=$ 5 $\times$ 10$^4$
MC steps. 
The results described below are insensitive to this cooling rate in the
sense that values of the freezing temperature $T^*$, for example, do not
change significantly even when the cooling rate is set faster by several
times.

In the present paper we examine a freezing parameter, ${\cal S}$,
in order to study freezing characteristics of the dipolar plate.  
It is defined as
\begin{eqnarray}
 {\cal S} = \frac{1}{N} \sum_i [| \langle\vec{S}_i \rangle|]_{\rm r},\label{freezingS}
\end{eqnarray}
where $N$ denotes the number of sites, $\langle ... \rangle$ thermal
average over the MC steps $\tau$, and $[...]_r$ average
over cooling runs. 
This parameter represents the frozen degree of spins and is important as
an order parameter of dipolar systems. 
In the present work, ${\cal S}$ is measured during cooling processes for
the dipolar hexagonal plate whose inter-layer distance $l$ is set from
$l_0/3$ to $l_0$, with $l_0$ being the inter-layer distance of the FCC
structure. 

\section{Freezing characteristics}\label{FPS}

\begin{figure} [t]
\begin{center}
\includegraphics[width=1.0\linewidth]{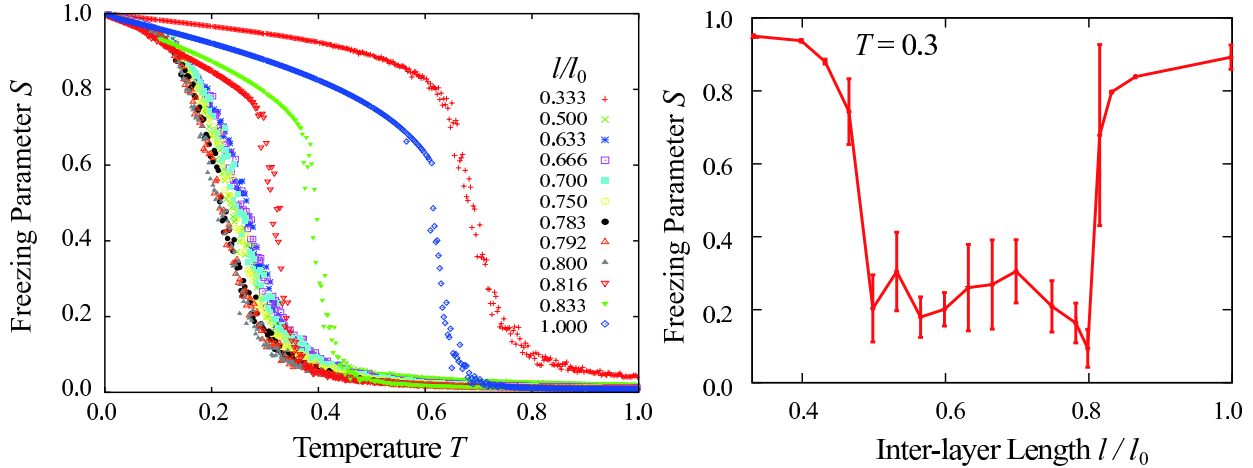}
\end{center}
\caption{(a) The temperature dependence of the freezing parameter for each $l$.\ (b) The freezing parameter as a function of $l$ at $T = 0.3$.}
\label{FP}
\end{figure}

In this section we discuss freezing characteristics based on the
freezing parameter whose temperature dependence is shown in 
Fig.~\ref{FP}(a). 
Data points with each symbol represent the values of ${\cal S}$ for each
fixed inter-layer distance, $l$. 
At $T \simeq T^*$, ${\cal S}$ for each $l$ starts to increases rapidly
and specific heat (not shown) exhibits a peak. 
Let us examine the behavior of $T^*$ in order to clarify trends of the
freezing characteristics (see also Fig.~\ref{SPD}).
At $l = l_0$, $T^*$ is nearly equal to 0.65. 
It steadily decreases as $l$ decreases from $l_0$.
When $l$ reaches nearly at 0.8$l_0$, $T^*$ reaches at its lowest value, 
which is nearly equal to 0.2.
As $l$ further deceases up to around 0.7$l_0$, $T^*$ slowly increases. 
After $l$ passes 0.7$l_0$, $T^*$ stays nearly constant up to $l$ of about
0.5$l_0$ where $T^*$ starts to increase rapidly.
When $l$ is further decreased, $T^*$ increases continuously.  

From the above observation of Fig.~\ref{FP}(a), we can expect at least
the following three ranges of $l$ in which the freezing characteristics
of the system differs significantly; $0.8l_0 \lesssim l$, 
$0.5l_0 \lesssim l \lesssim 0.8l_0$, and $l \lesssim 0.5l_0$.
Actually, at the corresponding temperatures such as at $T = 0.3$, we
observe the spin melting and refreezing respectively at $l \sim 0.8l_0$
and  $l \sim 0.5l_0$ driven by the uniaxial compression as shown in
Fig.~\ref{FP}(b).  
These spin melting and refreezing are related to changes of the
corresponding ground states as shown in Fig.~\ref{SPD} and discussed in
detail in the next section.

\section{Ground state}

\begin{figure}[t]
\begin{center}
\includegraphics[width=0.75\linewidth]{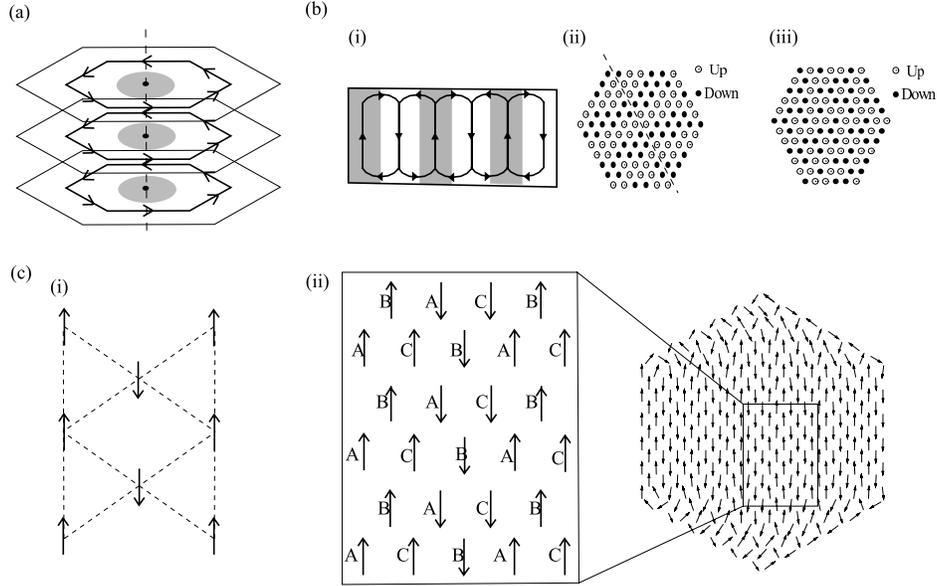}
\end{center}
\caption{The ground state for 
(a) $0.8l_0 \lesssim l$ (in-plane vortex), (b) $0.5l_0 \lesssim l \lesssim 0.8l_0$  (convex flow), and (c) $l \lesssim 0.5l_0$ (c) (in-plane-af-FMC). In each schematic picture, spin configurations are shown by arrows. The symbols A, B, and C in (c-ii) denote sites on the top, intermediate and bottom layers, respectively.}\label{GS}
\end{figure}

Let us first discuss the spin patterns of the ground state (or the state
reached by the present analysis at the lowest temperatures) 
of the plates with $l \gtrsim 0.8l_0$.
The one of $l = l_0$ is shown in Fig.~\ref{GS}(a). 
Spins are confined in the $x$-$y$ plane and they form a single vortex.
Such an in-plane vortex state is often observed in 2d dipolar
systems~\cite{Belobrov,Vedmedenko}.  
In the present dipolar hexagonal plate 
the vortex state consists of six ferromagnetic domains extended from all
the edges of the plate.
In the cooling process to this ground state 
the from-Edge-to-Interior freezing~\cite{KSKTT} is observed. 
Both the energetics of the ground state and the origin of this freezing
characteristics have been discussed in our previous paper~\cite{KSKTT2}.
In a whole range of $0.8l_0 \lesssim l$ this vortex state is observed at
lowest temperatures. 
When $l$ becomes close to $0.8l_0$, however, the in-plane vortex state
are gradually broken from around the center of every hexagonal layer, 
which is indicated by the gray region in Fig.~\ref{GS}(a). 
In this region spins possess an out-of-plane component. 

The ground state of the dipolar plate with 
$0.5l_0 \lesssim l \lesssim 0.8l_0$ is found to be a convex-flow
state schematically shown in Fig.~\ref{GS}(b-i). 
The $z$-component pattern of spins on the middle layer of the plate with
$l = 0.7l_0$ is shown in Fig.~\ref{GS}(b-ii), where a stripe pattern of
ferromagnetic orders is clearly seen.
In fact the absolute magnitudes of these $z$-components are close to
unity. 
Spins on the first (top) and third (bottom) layers lie almost within the
layers and align to connect smoothly with the neighboring ferromagnetic 
orders, yielding a convex-flow pattern as a whole.
At $l \simeq 0.5l_0$, the stripe pattern is destabilized to the one
shown in Fig.~\ref{GS}(b-iii). 
We may call the latter as a randomly-distributed convex-flow state. 
Whether the range of $l$, where this state appears, is finite or not has
not been confirmed, but so far it has not been observed at the
other end of the convex-flow state, i.e., at $l \simeq 0.8l_0$.

For further compressed systems such as the one with $l = l_0/3$,
an in-plane state again appears as shown in Fig.~\ref{GS}(c).
Each on-layer pattern of this state, however, has not the
vortex structure but the anti-ferromagnetic ferromagnetic-chain (af-FMC)
order shown in Fig.~\ref{GS}(c-i); spins on one lattice axis align 
ferromagnetically directed to the axis, and this ferromagnetic chain
order alternates in the direction perpendicular to the axis.
It is similar to one of the (continuous) ground states of the bulk
dipolar system on the 2d square lattice~\cite{DBell}. 
In the present hexagonal plate, the on-layer af-FMC orders align
ferromagnetically in the $z$-direction as shown in
Fig.~\ref{GS}(c-ii). 
Spins on two sides of the plate direct within the side plane keeping the
af-FMC order with the interior spins, while those on the other four sides 
align to connect continuously between the af-FMC pattern on the top
layer and that on the bottom layer. 
The present in-plane-af-FMC ground state is confirmed in the range 
$0.33l_0 \le l \le 0.46l_0$ at least.
We expect one more change of the ground state at a further 
shorter $l$ than $0.33l_0$ as shown in Fig.~\ref{SPD}, because at $l = 0$ 
the system is a 2d triangular lattice layer whose density of spins
is tripled as compared with the original one and so the vortex state
is naturally expected. 


\section{Discussion}

In our recent work~\cite{KSKTT2} the same dipolar hexagonal systems but
with the numbers of layers, $N_{\rm L}$, up to 15 and with 
$l=l_0,\ l_0/{\sqrt 2}$ were studied. 
We observed the in-plane-vortex ground state in systems with 
$l=l_0\ (l_0/{\sqrt 2})$ for all $N_{\rm L}$'s ($N_{\rm L} \le 2$), and
the convex-flow ground state for $N_{\rm L} \ge 3$ in systems
with $l=l_0/{\sqrt 2}$. 
In a bulk system with $l=l_0$, i.e., the FCC lattice, the ferromagnetic
order has the global $O(3)$ symmetry~\cite{LT}.
Therefore, when the hexagonal boundary is introduced, there appear the
in-plane-vortex ground state which gains the boundary-induced
anisotropic energy including that from all edges.
When the FCC structure is compressed in the $z$-direction, the
magnetization prefers to align to this direction, and the competition
with the boundary-induced anisotropic energy takes place.
An important finding of the present work is that the critical distance,
$l_{\rm cr}$, at which the ground state changes from the in-plane-vortex
to the convex-flow ones, is given as $l_{\rm cr} \simeq 0.8l_0$ for
$N_{\rm L}=3$.   
More interestingly, the spin melting phenomenon is observed at low
temperatures, such as $T \sim 0.3$, as $l$ decreases around 
$l_{\rm cr}$, correspondingly to the ground state transition.
Furthermore, when $l$ is further decreased, the spin refreezing    
phenomenon is found. It corresponds to the ground state change to the
in-plane-af-FMC one. 
The occurrence of the latter is certainly related to the fact that the
present system with $l=l_0/2$ is just a simple cubic lattice whose one
of the bulk ground states is the af-FMC one~\cite{LT}, though details of
the shape effect are not completely understood yet.

The results so far described are obtained for the hexagonal dipolar
plate by varying $l$ but with $N_L\ (=3)$ and a size of the plate fixed. 
Although we have not completed MC simulations with systematic change
of the other parameters, nor have analyzed effects of the boundary
roughness, the present results are enough to demonstrate that the
dipolar systems of a finite size have a vast variety of magnetic
phenomena, which are hardly thought of for ordinary spin systems only
with simple exchange-type interactions. 


The present work is supported by the NAREGI Nanoscience Project of the
Ministry of Education, Culture, Sports, Science, and Technology. The
numerical simulations have been partially performed also by using the 
facilities at the Supercomputer Center, Institute for Solid State
Physics, the University of Tokyo.


\section*{References}
\begin{thebibliography}{10}
\bibitem{Martin}J~J~Martin, J~Nogues, K~Liu, J~L~Vicent and I~Schuller
        2003 {\it J. Mag. Mag. Matel.} {\bf 256} 449.
\bibitem{Garel} T~Garel and S~Doniach 1982 {\it Phys. Rev. B} {\bf 26} 325 
\bibitem{Igresias} J~R~Iglesias, S~Gom\c{c}alves A~Nagel and M~Kiwi 2002 {\it Phys. Rev. B} {\bf 65} 064447
\bibitem{Jagla} E~A~Jagla 2004 {\it Phys. Rev. E} {\bf 70} 046204
\bibitem{Belobrov} P~I~Belobrov, V~A~Voevodin and V~A~Ignatchenko 1985 {\it Sov. Phys. JETP} {\bf 61} 522
\bibitem{Vedmedenko} E~I~Vedmedenko, A~Ghazali and J-C~S.Le\`vy 1999 {\it Phys. Rev. B} {\bf 59}, 3329
\bibitem{KSKTT} K~Matsushita, R~Sugano, A~Kuroda, Y~Tomita, H~Takayama 2005 {\it J. Phys. Soc. Jpn.} {\bf 74} 2651
\bibitem{KSKTT2} K~Matsushita, R~Sugano, A~Kuroda, Y~Tomita, H~Takayama to be published
\bibitem{DBell} K~De'Bell, A~B~MacIsaac, I~N~Booth and J~P~Whitehead 1997
Phys. Rev. B {\bf 55} 15108
\bibitem{LT} J.~M.~Luttinger and L.~Tisza 1946 Phys. Rev. {\bf 70} 954
\end{thebibliography}
\end{document}